\documentclass[aps,pra,showpacs,floatfix,superscriptaddress,twocolumn]{revtex4}

\usepackage{amsmath}
\usepackage{amssymb}
\usepackage{dsfont}
\usepackage{graphicx}
\usepackage{color}
\usepackage{mathrsfs}
\usepackage{mathtools}
\usepackage{braket}
\usepackage{epstopdf}

\def\vec#1{{\boldsymbol{#1}}}
\def\<{\langle}
\def\>{\rangle}

\def\Naturals{\mathbb N}

\def\Tr{\operatorname{Tr}}

\newtheorem{lemma}{Lemma}[section]

\newtheorem{theorem}[lemma]{Theorem}

\begin{document}
\title{Dominant couplings in qubit networks with controlled interactions}

\author{{\qquad\qquad\qquad\qquad\qquad\qquad\qquad\qquad\qquad\ \ J. Mary\v ska, J. Novotn\'y, I. Jex}\newline
\emph{
Department of Physics, Faculty of Nuclear Sciences and Physical Engineering,
Czech Technical University in Prague, B\v rehov\'a 7,
115 19 Praha 1 - Star\'e M\v{e}sto, Czech Republic \\
}}

\date{\today}

\begin{abstract}
Systems evolving under the influence of competing two-body and three-body interactions, are of particular interest in exploring the stability of equilibrium states of strongly interacting many-body system. We
present a solvable model based on qubit networks, which allows us to investigate the intricate influence of these couplings on the possible asymptotic equilibrium states. We study the asymptotic evolution of finite qubit networks under two and three-qubit interactions. As representatives of three-qubit interactions we choose controlled unitary interactions
(cu-interactions) with one and two control qubits. It is shown that networks with purely three-qubit interactions exhibit different asymptotic dynamics depending on whether we deal with interactions controlled by one or two qubits. However, when we allow three-qubit interactions next to two-qubit interactions, the asymptotics is dictated by two-qubit interactions only. Finally, we prove that the simultaneous presence of two types of three-qubit interactions results in the asymptotic dynamics characteristic for two-qubit cu-interactions.
\end{abstract}

\maketitle

\section{Introduction}
\label{part_I} The evolution of any physical system is governed by
the interactions between the constituting parts and the external
conditions. Especially in statistical physics the understanding of
the form of the internal interactions plays a crucial role in the
description of the system. Usually we limit ourselves to two-body
interactions and neglect higher order (many-body) interactions. The
ignored interactions have usually properties which differ from two-body interactions and induce different behavior of the system.
In particular, states which are stable under two-body interaction may become unstable under the influence of additional three-body interaction as observable in undercooled gases, for example. Hence there is certainly interest in situations in which competing types of interactions can be solved analytically and allow to specify the role of the additional contributions. We attack this
problem by using the framework of quantum networks and study the
dynamics of arbitrary sized network with competing two-body and three-body interactions. The elementary interactions are given in terms of
quantum gates.

Qubit systems play an important role in quantum theory.
Beside their importance as nontrivial representations of many
physical systems they are of significant use in quantum information
theory \cite{nielsen, barnett, schumacher}.
Coupled qubits equipped with a special type of interaction can be
used in quantum computing to process and manipulate information. The
mutual couplings do not necessarily involve only particular qubits,
they can be also a result of an interaction with an additional
auxiliary system. Qubit systems of this kind are generalizations
of classical networks \cite{spackove, cvrcci}. They are
therefore called qubit networks and they can be represented
by graph structures. Nodes are qubits (equipped with two base states $\ket{0}$ and $\ket{1}$),
edges (directed or undirected) represent interactions between nodes.
Qubit networks are a part of a broader family of quantum networks.
Because of their importance, quantum networks received a lot of attention in
recent years \cite{qinternet, qnet1, qnet2}. They offer a wide variety of utilizations, e.g.
establishing entanglement for various applications \cite{cirac1, cirac2, entest}, modelling of quantum decoherence \cite{decoherence1, decoherence2} or reaching quantum consensus \cite{qnet3,qnet4,qnet5}.

We often find ourselves in situation, in which we are unable to
control the type of interaction or the strength of the coupling
between qubits. As a result, classical randomness must be introduced into
such a qubit network. A convenient way how to describe such a dynamics are Markov chains,
i.e. sequences of random (quantum) operations.
In a significant number of cases the situation can be mathematically described by a superoperator
which represents a random unitary operation (RUO) \cite{novotny1}.
This superoperator is equipped with a probability distribution over
all possible qubit couplings that can be realized in a single
application of the given RUO. The resulting time evolution is then given
by an iteration process of repeated application of the given RUO on
the density matrix describing the state of the system in the
previous step.

Recently, quite a general class of two-qubit interactions has been
considered. The interaction between qubits has been chosen as a member of a one-parameter family of two-qubit cu-interactions (i.e. controlled unitary interactions) \cite{decoherence1, novotny2}.
The evolution of qubit networks with such interaction was calculated in the asymptotic
regime, i.e. after a sufficiently large number of interactions. The
description had taken into account the topology of the given
qubit network, i.e. realizable couplings between qubits. Such qubit
networks can be used to model a number of physically relevant and
interesting situations, e.g. they can be used as a simple
model of particle collisions in a rarefied gas or as a class of
one-parameter decoherence models \cite{decoherence1}. Since the asymptotic
state is independent of the non-zero values of the probability
distribution over the qubit couplings \cite{novotny1}, it is of interest to find out,
how does the asymptotic state change, if we enrich the
dynamics of the network by adding higher-order interactions,
e.g. three-qubit interactions as the simplest generalization. One is
also interested in properties of interactions of higher orders on
their own, i.e. without the presence of two-qubit interactions, as the
results obtained previously cannot be extended to more general
cases in a straightforward way. Higher order interactions
can be considered a step towards more complicated interaction structures
which are not decomposible into simple (random) two-qubit interaction.

In the following, we investigate the effect of three-qubit cu-interactions
on the qubit network equipped with two-qubit cu-interactions.
The considered three-qubit cu-interactions have form of controlled unitary three-qubit
interactions with one or two control qubits. Both cases are
studied in detail for a class of interaction topologies represented by so-called base graphs.
First, we generalize the concept of an interaction graph \cite{novotny2}
(used to specify realizable two-qubit couplings) to the
case of three-qubit couplings. Next, we derive the form of asymptotic
states of three-qubit cu-interactions for certain family of interaction topologies.
This enables us to study properties of qubit networks
equipped with purely three-qubit cu-interactions as well as properties
of qubit networks equipped with both two and three-qubit cu-interactions.

The paper is structured as follows. Section \ref{part_II} provides
the theoretical background of basic concepts of graphs and hypergraphs,
which are used through the rest of the paper. Section \ref{part_III} summarizes basic properties of RUOs which are required for finding the asymptotic state (for more details see \cite{novotny1}). Results relevant to two-qubit cu-interactions are given in section \ref{part_IV}. Rigorous definitions of three-qubit cu-interactions
and corresponding graph structures used through the paper are given in section \ref{part_V} as well as the derivation of corresponding asymptotic states.
The relation between two and three-qubit cu-interactions is studied in
section \ref{part_VI}. Section \ref{part_VII} presents additional properties of
three-qubit cu-interactions. Summary and outlook of the paper are left to section \ref{part_VIII}.
Main steps of proofs of theorems presented in section \ref{part_V} are given in appendix.

\section{Basic definitions in graph and hypergraph theory}
\label{part_II}
Let us first introduce graph structures which are
convenient for the description of multi-qubit cu-interactions.
A natural and often used way to inscribe a topology of two-qubit
cu-interactions in qubit networks constitute directed graphs.
A directed graph is defined as an ordered pair $G=(V,E)$,
where $V=\{v_1,\dots v_N\}$ is the set of vertices and
$E=\{e_1\dots e_m\}$ is the set of edges. Each edge can be expressed
as a pair of vertices, which are called the tail and the head of the given edge,
 hence $e_i=(v_{i_t},v_{i_h})$. A sequence of vertices
$(v_{i_1},\dots,v_{i_k})$ such that $(v_{i_j},v_{i_{j+1}})$
forms an edge in G is called a directed path connecting vertices $v_{i_1}$
and $v_{i_k}$. These vertices are called the beginning and
the end of the path respectively. If every two vertices of the directed graph $G$
are connected by a path in both directions $G$ is said to be strongly connected.

A hypergraph \cite{hypergraphs} is a direct generalization of a graph
in which we do not restrict relations between vertices to be
binary ones. Instead the directed hypergraph is defined as an ordered
pair $H=(V,E)$ where $E$ is an arbitrary subset of the set constituted by all pairs of
disjoint subsets $X\subset V$. Thus every $e\in E$ can be written as $e=(X,Y)$
with $X,Y\subset V$, $X,Y\neq\emptyset$ and $X\cap Y=\emptyset$. If $u\in X$, then $u$ is
called the tail of hyperedge $e$, if $v\in Y$, then $v$ is called the head of
hyperedge $e$. A directed path in a hypergraph is defined as a sequence
of vertices $(v_{i_1},\dots,v_{i_k})$ such that for every $j$ there is
$(X,Y)\in E$ such that $v_j\in X$ and $v_{j+1}\in Y$. A directed hypergraph
H is called strongly connected if every pair of vertices can be connected
by a directed path in both directions.

There exist special types of hyperedges which we find particularly useful
for describing the topology of multi-qubit cu-interactions.
Let $e=(X,Y)$ be a hyperedge. If $|X|=1$, $e$ is called a F-arc,
if $|Y|=1$, $e$ is called a B-arc. A hypergraph $H$ in which all
hyperedges are F-arcs is called a F-graph, a hypergraph $H$ in
which all hyperedges are B-arcs is called a B-graph. Ordinary graphs belong to both these sets.

Natural partial ordering on hypergraphs is defined in the following way. Let $G=(V,E)$ and $H=(U,F)$ be two hypergraphs. We say that $G$ is a subgraph of $H$, i.e. $G\subset H$, if $G$ and $H$ are defined on the same set of vertices and $e\in E$ implies $e\in F$.

\section{Attractor method for dynamics generated by RUO}
\label{part_III}
In this paper we study quantum systems undergoing discrete dynamics whose one step of evolution is described by a random unitary operation (RUO). General properties of a random unitary evolution with special emphasis on its asymptotic regime was extensively studied in \cite{novotny1}. We briefly summarize main findings important for our considerations to follow.

Assume a quantum system associated with a finite dimensional Hilbert space $\mathscr{H}$ and let us denote $\mathcal{B}(\mathscr{H})$ the Hilbert space of all operators acting on the Hilbert space $\mathscr{H}$ (equipped with Hilbert-Schmidt scalar product \cite{nielsen}). One step of evolution is given by the RUO $\Phi:\mathcal{B}(\mathscr{H})\rightarrow\mathcal{B}(\mathscr{H})$
whose action on the system initially prepared in a general mixed state $\rho$
can be written in the form
\begin{equation}
\label{def_RUO}
\Phi(\rho)=\sum_{i=1}^n p_iU_i\rho U_i^{\dagger}
\end{equation}
with unitary operators $U_i$ and probability distribution $\left\{p_i \right\}_{i=1}^n$. RUO $\Phi$ is a quantum operation with Kraus operators defined by $K_i=\sqrt{p_i}U_i$. It belongs to the class of trace-preserving unital quantum operations leaving the maximally mixed state invariant. From a physical point of view, RUO takes into account our lack of information which particular unitary evolution the system undergoes and incorporates in an incoherent manner all exclusive unitary paths of evolution represented by different unitary operators $U_i$ properly weighted with probabilities associated with all these paths.

The evolution of the system results from repeated applications of RUO $\Phi$.
Starting from the initial state $\rho(0)$, the $n$-th step of the iterated dynamics reads $\rho(n)=\Phi(\rho(n-1))$. In general, the operator $\Phi$ is neither a hermitian nor normal linear map and consequently a diagonalization of RUO $\Phi$ in some orthonormal basis is not guaranteed. Fortunately, the latter does not apply to the asymptotic part of evolution and one can exploit the fact that the asymptotic
regime of the evolution takes place in the so-called attractor space $\text{Atr}(\Phi)\subset\mathcal{B}(\mathscr{H})$ constructed as
\begin{equation}
\text{Atr}(\Phi)=\bigoplus_{\lambda\in\sigma_{|1|}}\text{Ker}(\Phi-\lambda I).
\end{equation}
The attractor spectrum $\sigma_{|1|}$ denotes the set of all eigenvalues $\lambda$ of $\Phi$
with $|\lambda|=1$. For a given $\lambda\in\sigma_{|1|}$ the
corresponding kernel $\text{Ker}(\Phi-\lambda I)$ is formed by
the solution of attractor equations
\begin{equation}
\label{eq_attractor_general}
U_i X_{\lambda,j}U_i^{\dagger}=\lambda X_{\lambda,j}\quad \forall i.
\end{equation}
Any state in the asymptotic regime takes the form
\begin{equation}
\label{eq_asymptotic_evolution_gen}
\rho(n\gg 1)=\sum_{\lambda\in\sigma_{|1|},i=1}^{D_{\lambda}}\lambda^n\text{Tr}\left[\rho(0) X_{\lambda,i}^{\dagger}\right] X_{\lambda,i},
\end{equation}
where
$D_{\lambda}=\text{dim}[\text{Ker}(\Phi-\lambda I)]$
and
$\{X_{\lambda,i}|i\in\{1,\dots,D_{\lambda}\}\}$
forms an orthonormal basis of the subspace $\text{Ker}(\Phi-\lambda I)$ with respect to the Hilbert-Schmidt scalar product.
Apparently, attractors solving equations (\ref{eq_attractor_general}) are not affected by the particular
form of the probability distribution $\{p_i\}$ provided that $p_i\neq 0$.
As a direct consequence, the asymptotics of given RUO is independent on particular values $p_i$ as long as they are different from zero.

At this point we have to stress that attractors are not, in general, density operators, i.e. they do not represent states. This can cause difficulties in the analysis of asymptotic dynamics like fixed points, decoherence-free subspaces. To overcome this obstacle one may employ the so-called pure-state method \cite{percolation}. Without going into details we recapitulate its main points. Let us denote $\left\{ |\phi_{\alpha,j_{\alpha}}\>\right\}$ the orthonormal basis of common eigenstates of unitaries $\{U_i\}$, i.e.
\begin{equation}
U_i |\phi_{\alpha,j_{\alpha}}\> = \alpha |\phi_{\alpha,j_{\alpha}}\> \qquad \forall i,
\label{common_pure_eigenstates}
\end{equation}
where index $j_{\alpha}$ takes into account the degeneracy of common eigenvalue $\alpha$. Any matrix from the span of $\{|\phi_{\alpha,j_{\alpha}}\> \< \phi_{\beta,j_{\beta}}|\}$ with fixed product $\alpha \overline{\beta} =\lambda$, i.e.
\begin{equation}
X=\sum_{\alpha \overline{\beta}=\lambda,j_{\alpha},j_{\beta}} A^{\alpha,j_{\alpha}}_{\beta,j_{\beta}}|\phi_{\alpha,j_{\alpha}}\> \< \phi_{\beta,j_{\beta}}|,
\label{p_attractors}
\end{equation}
satisfies equations
\begin{equation}
U_i X U_j^{\dagger} = \lambda X \qquad \forall i,j
\label{eq_p_attractor_gen}
\end{equation}
and consequently belongs to the subspace of attractors corresponding to the eigenvalue $\lambda$ of RUO $\Phi$. On the other hand, any operator satisfying (\ref{eq_p_attractor_gen}) can be decomposed into common eigenvectors as (\ref{p_attractors}). Attractors which can be constructed from common eigenvectors are called p-attractors. As they satisfy more restricting set of equations (\ref{eq_p_attractor_gen}) they do not constitute the whole attractor space. In particular, the identity operator is an attractor but not a p-attractor (except the trivial case of an unitary evolution). The space of attractors always contains, as the minimal subspace, the span of p-attractors and the identity operator. Surprisingly, in some nontrivial cases this minimal subspace forms the whole attractor set and the asymptotic dynamics can be analyzed easier. Indeed, assume the orthogonal projection $\mathcal{P}$ onto the subspace of common eigenstates of unitaries $\{U_i\}$. Let $\tilde{\mathcal{P}}$ be its orthogonal complement projection satisfying $\mathcal{P} + \tilde{\mathcal{P}}=I$. Both projections are fixed points of quantum operation (\ref{def_RUO}), they both reduce the random unitary operation (\ref{def_RUO}) and the asymptotic dynamics of the initial state $\rho(0)$ can be written as
\begin{equation}
\label{eq_unitary_evolution}
\rho(n \gg 1) = U_i^n \mathcal{P} \rho(0) \mathcal{P} \left(U_j^{\dagger}\right)^n + \frac{\Tr\left[ \rho(0) \tilde{\mathcal{P}}\right]}{\Tr\left[ \tilde{\mathcal{P}}\right]} \tilde{\mathcal{P}},
\end{equation}
for any pair of indices $i,j$. In this case the asymptotic evolution can be understood as an incoherent mixture of the unitary dynamics inside the subspace of common eigenstates and the maximally mixed state living on the orthogonal complement of this subspace. We often encounter the special case $\sigma_{|1|}=\{1\}$, which further simplifies the asymptotic evolution given by (\ref{eq_unitary_evolution}) to the form
\begin{equation}
\label{eq_evolution_stationary}
\rho_{\infty}=\mathcal{P} \rho(0) \mathcal{P} + \frac{\Tr\left[ \rho(0) \tilde{\mathcal{P}}\right]}{\Tr\left[ \tilde{\mathcal{P}}\right]} \tilde{\mathcal{P}}.
\end{equation}
In such a situation any initial quantum state $\rho(0)$ evolves towards the stationary state (\ref{eq_evolution_stationary}).

Let us finally point out a crucial property of general attractor equations (\ref{eq_attractor_general}) which we extensively use throughout the paper. We formulate this property in connection with partial ordering on hypergraphs. In order to proceed consider a hypergraph $H=(V,E)$ and assume that for any hyperedge $e \in E$ there is an unitary operation $U_e$ and nonzero probability $p_e$ such that $\sum_{e \in E} p_e =1$. RUO associated with the given hypergraph is defined as
\begin{equation}
\label{def_RUO_hypergraph}
\Phi_{G}(\rho)= \sum_{e\in E}p_eU_e \rho U_e^{\dagger}.
\end{equation}
If two hypergraphs $G$ and $H$ satisfy $G \subset H$ then attractor spaces of associated RUOs follow relation $\text{Atr}(\Phi_H)\subset\text{Atr}(\Phi_G)$. This is an immediate consequence of the fact that the hypergraph $H$ implies more
constraints than the hypergraph $G$. These relations imply the existence of the set of so-called base attractors. Indeed, assume a subset of hypergraphs. Any hypergraph from this set determines corresponding RUO (\ref{def_RUO_hypergraph}). The subspace of base attractors is identified as the maximal set of attractors included in attractor spaces of all these associated RUOs. This subspace always contains the identity operator. Similarly, we define the set of base hypergraphs whose associated attractor spaces are formed from base attractors solely. In other words, the attractor subspace of all RUOs associated with base graphs is constituted from base attractors.

It is important to stress that the definition of base attractors as well as the definition of base graphs is closely linked to the chosen set of hypergraphs. In particular, in this paper we always consider the set of graphs or a certain subset of F-graphs.

\section{Asymptotic dynamics of two-qubit cu-interactions}
\label{part_IV}
Before discussing three-qubit cu-interactions we shortly review known results on
asymptotic dynamics generated by two-qubit cu-interactions which have been recently studied \cite{novotny1, novotny2, decoherence1}. For this purpose
let us consider a qubit network consisting of $N$ qubits in which
pairs of qubits are randomly coupled to each other via certain
transformations according to the prescribed probability distribution.
Information about qubit couplings can be encoded into a directed graph
$G=(V,E)$ which is called the interaction graph. Each vertex of
this graph represents a single qubit, e.g. the vertex $i$ represents
the qubit $i$. Edges of the interaction graph then represent
couplings between qubits, e.g. the edge $e=(i,j)$ represents the
coupling between qubits $i$ and $j$. This edge is present in
the interaction graph if only if the corresponding
probability of realization of given coupling is nonzero.
The interaction graph defined above
gives us sufficient amount of information about the given RUO
as the precise form of the probability distribution $\{p_i\}$
is irrelevant for the asymptotic dynamics generated by RUO.

In view of equations (\ref{eq_attractor_general}), the topology of the interaction graph determines the attractor space associated with a given two-qubit cu-interaction and consequently the asymptotic evolution of any initial state (\ref{eq_asymptotic_evolution_gen}). Assuming all interaction graphs the main task is to identify their corresponding subspace of base attractors and the set of base graphs (section \ref{part_III}). Previous work was focused on a one-parameter family of two-qubit cu-interactions including the controlled-NOT transformation as its special case. In the computational basis the coupling between qubits $i$ and $j$ reads
\begin{equation}
U_{ij}^{(\phi)}=\ket{0_i}{\bra{0_i}}\otimes I_j+\ket{1_i}{\bra{1_i}}\otimes U_j^{(\phi)},
\end{equation}
\noindent where
\begin{equation}
\label{def_interaction}
U_j^{(\phi)}=\cos(\phi) Z_j+\sin(\phi) X_j
\end{equation}
with the identity operator $I_j$ and Pauli operators $X_j$ and $Z_j$ acting on the $j$-th qubit \cite{nielsen}. The parameter $\phi$ is chosen from
the interval $(0,\pi)$. As these couplings have
the form of asymmetric control transformations, the interaction graph becomes
a directed graph in which the transformation
$U_{ij}^{(\phi)}$ is represented by the directed edge $e$
with the vertex $i$ as the tail of $e$ and the vertex $j$
as the head of $e$. Once the interaction graph is given the corresponding RUO $\Phi^{(2)}$ describing one step of evolution is defined as
\begin{equation}
\label{eq_ruo_two_qubit}
\Phi^{(2)}(\rho)=\sum_{e\in E}p_eU_e^{(\phi)}\rho \left(U_e^{(\phi)}\right)^{\dagger}.
\end{equation}
The asymptotic behavior of dynamics generated by (\ref{eq_ruo_two_qubit}) can be summarized as follows (for details see \cite{novotny2}). It was shown that base graphs are formed by all strongly
connected graphs. The space of common eigenvectors for strongly connected networks is two-dimensional and its basis reads
\begin{eqnarray}
\ket{\textbf{0}_N}&=&\ket{0}^{\otimes N}, \nonumber \\
\ket{\varphi_N^+}&=&\left(\cos\frac{\phi}{2}\ket{0}+\sin\frac{\phi}{2}\ket{1}\right)^{\otimes N}. \nonumber
\end{eqnarray}
Both these common eigenvectors correspond to the eigenvalue $\alpha=1$. According to section \ref{part_III}, p-attractors constitute $4$-dimensional space and in order to complete the attractor space one has to add the identity operator. Only for $N=2$ qubits there is also one extra non-p-attractor corresponding to eigenvalue $\lambda=-1$, namely the operator $X_{-1}=\cos(\phi/2)(|01\>\<11| -|10\>\<11|) +\sin(\phi/2)|01\>\<10| ++\ h.c. $
Apart of the special case $N=2$, the asymptotic dynamics is stationary and takes the form (\ref{eq_evolution_stationary}).

The main goal of this paper is to analyze how an interplay among two-qubit and three-qubit cu-interactions reshapes the explicit form of base attractors and base graphs and determines the resulting asymptotic dynamics. To gain insight into this issue one has to first reveal the structure of base attractors and base graphs for purely three-qubit cu-interactions.

\section{Asymptotic dynamics of three-qubit cu-interactions}
\label{part_V}
The three-qubit cu-interaction does not reduce simply
to the two-qubit cu-interaction and hence this situation requires
separate attention. For this reason we examine three-qubit
cu-interactions of two different forms by changing the number of
control qubits. The main purpose is twofold. First, we want to
compare properties of asymptotic states of two-qubit cu-interactions
with properties of asymptotic states of three-qubit cu-interactions.
Second, we intend to reveal the relationship between base attractor
spaces of two-qubit and three-qubit cu-interaction networks. In particular,
it is an important question, whether the base attractor space of two-qubit cu-interactions is a subspace of the base attractor space of three-qubit cu-interactions. In the case of a positive answer, the set of attractor equations (\ref{eq_attractor_general}) implies that by enabling three-qubit cu-interactions next to two-qubit cu-interactions, one does not change the asymptotic state at all as long as the interaction graph corresponding to the two-qubit cu-interaction remains strongly connected.

\subsection{Controlled unitary three-qubit transformation with one control qubit} \label{part_Va}
\begin{figure}[t!]
\centering
\includegraphics{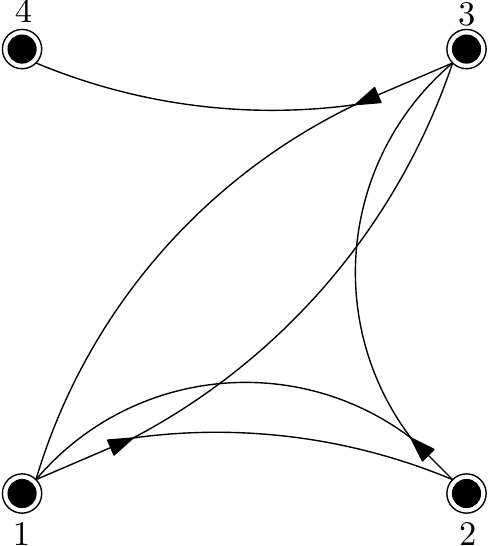}
\caption{An example of an interaction $\text{F}_1$-graph $G$ corresponding to the RUO $\Phi_G^{(3,1)}$ acting on $4$ qubits. It involves three unitary operations corresponding to three hyperedges $(1,23)$, $(3,14)$, and $(2,13)$. This $\text{F}_1$-graph is not strongly connected and thus does not belong to the set of base graphs.}
\label{fig_controlled_1_2}
\end{figure}

Let us consider a qubit network in which qubits are
coupled to each other by three-qubit transformations which
in the computational basis take the form
\begin{equation}
\label{def_controlled_1_2}
U_{i,jk}^{(\phi)}=\ket{0_i}{\bra{0_i}}\otimes I_{jk}+\ket{1_i}{\bra{1_i}}\otimes U_j^{(\phi)}\otimes U_k^{(\phi)}
\end{equation}
with $U_j^{(\phi)}$ defined by (\ref{def_interaction}).
The set of all assumed three-qubit controlled interactions (\ref{def_controlled_1_2}) is prescribed by an interaction $\text{F}_1$-graph $G=(V,E)$ (section \ref{part_III}). The set of vertices $V=\{1,\dots,N\}$ corresponds to the set of qubits and F-arcs $(X,Y)\in E$ with $|Y|=2$ represent possible three-qubit cu-interactions. In particular, if $p_{i,jk}>0$ then the vertices $i, j$ and $k$ are connected in $G$ by a F-arc $e\in E$ whose head are qubits $j$ and $k$ and whose tail is the qubit $i$. Hyperedges of any $\text{F}_1$-graph thus consist of two heads and one tail. An example of a network representing a topology of three-qubit cu-interactions with one control qubit is depicted in Fig.\ref{fig_controlled_1_2}.

Once an interaction $\text{F}_1$-graph $G=(V,E)$ is given the associated RUO describing one step of evolution reads
\begin{equation}
\label{def_RUO_1_2}
\Phi^{(3,1)}(\rho)=\sum_{e\in E}p_eU_e^{(\phi)}\rho \left(U_e^{(\phi)}\right)^{\dagger}.
\end{equation}
The eigenvalues of the transformation (\ref{def_controlled_1_2}) are $\alpha=\pm 1$ and thus the attractor spectrum $\sigma_{|1|}$ of RUO (\ref{def_RUO_1_2}) is a subset of  $\{-1,1\}$. The following theorem fully describes base $\text{F}_1$-graphs and their corresponding base attractors of RUOs (\ref{def_RUO_1_2}) associated with interaction $\text{F}_1$-graphs. The sketch of the proof is given in appendix.
\begin{theorem}
\label{theorem_1_2}
Base $\text{F}_1$-graphs of the considered
family of interactions are all strongly connected interaction
$\text{F}_1$-graphs $G=(V,E)$. The space of common eigenvectors is three-dimensional with the basis
\begin{eqnarray}
\ket{\normalfont{\textbf{0}}_N}&=&\ket{0}^{\otimes N}, \nonumber \\
\ket{\varphi_N^+}&=&\left(\cos\frac{\phi}{2}\ket{0}+\sin\frac{\phi}{2}\ket{1}\right)^{\otimes N}, \nonumber\\
\ket{\varphi_N^-}&=&\left(\sin\frac{\phi}{2}\ket{0}-\cos\frac{\phi}{2}\ket{1}\right)^{\otimes N} \nonumber.
\end{eqnarray}
All these common eigenvectors correspond to the eigenvalue $\alpha=1$. Base p-attractors form a 9-dimensional subspace of the base attractor space corresponding to the eigenvalue $\lambda=1$. In order to complete the attractor space, several cases need to be distinguished.
For $\phi\neq\frac{\pi}{2}$, the base attractor space consists of the base p-attractors and the identity operator. If $\phi=\frac{\pi}{2}$ and $N>3$ there are, besides p-attractors, two additional linearly independent attractors corresponding to eigenvalue $\lambda=1$
\begin{align*}
 I_{N,E}&=\sum_{z\in I(N)}\left(1+(-1)^{\tau(z)}\right)\ket{z}\bra{z},\\
 I_{N,O}&=\sum_{z\in I(N)}\left(1-(-1)^{\tau(z)}\right)\ket{z}\bra{z},
\end{align*}
\noindent with $I(N)$ denoting the set of all possible binary
$N$-tuples and $\tau(z)$ is the sum of the bit values of the N-qubit string z. Only in the special case $\phi=\frac{\pi}{2}$ and $N=3$ qubits
there is an additional attractor $ X_{-1}$ corresponding to the eigenvalue $\lambda=-1$ given by
\begin{align*}
 X_{-1}&=\frac{1}{6}(\ket{101}\bra{011}-\ket{110}\bra{011}+\ket{110}\bra{101})+h.c.\\
\end{align*}
\end{theorem}

Using theorem \ref{theorem_1_2}, one can write down asymptotic evolution generated by RUO (\ref{def_RUO_1_2}). In the case $N\geq 3\wedge\phi\neq\frac{\pi}{2}$, the initial quantum state $\rho(0)$
approaches the stationary state (\ref{eq_evolution_stationary}) with $\mathcal{P}^{(3,1)}$ being projector
on the space of common eigenvectors spanned by vectors $\ket{\textbf{0}_N}, \ket{\varphi_N^+}$, and $\ket{\varphi_N^-}$.

\subsection{Controlled unitary three-qubit transformation with two control qubits} \label{part_Vb}
In the second case we consider a qubit network
in which qubits are coupled to each other by three-qubit
transformations. The interaction (in the computational basis) takes the form
\begin{equation}
\label{def_RUO_2_1}
U_{ij,k}^{(\phi)}=\left( I_{ij}-\ket{1_i1_j}{\bra{1_i1_j}}\right)\otimes I_k+\ket{1_i1_j}{\bra{1_i1_j}}\otimes U_k^{(\phi)},
\end{equation}
with $U_k^{(\phi)}$ defined by (\ref{def_interaction}). The considered RUOs have the form
\begin{equation}
\label{def_RUO_2_1b}
\Phi^{(3,2)}(\rho)=\sum_{e\in E}p_eU_e^{(\phi)}\rho \left( U_e^{(\phi)}\right)^{\dagger}.
\end{equation}
The family of unitary transformations (\ref{def_RUO_2_1})
includes also the Toffoli gate. Similar to the previous case, $E$ is
the subset of the set of partially ordered triples $(i,j;k)=(j,i;k)$. The eigenvalues of the transformation $U_{ij,k}^{(\phi)}$ are again given by $\alpha=\pm 1$ and the attractor spectrum $\sigma_{|1|}$ thus fulfils the relation $\sigma_{|1|}\subset\{-1,1\}$.

B-graphs seem to be a natural choice to describe the topology of three-qubit cu-interactions (\ref{def_RUO_2_1}). However in order to fully describe the topology of base graphs
in the case of controlled unitary interactions with two control qubits, we must significantly change the graph representations of given interactions. The required graph representation should provide crucial information whether any pair of control qubits can "see" all other qubits via interaction with them. This interaction does not have to be the direct one, it can be mediated by other pairs of qubits. This "visibility" is mathematically represented by paths in given graph representation. For this reason the representation constructed analogically as the previously used representations (interaction graphs and interaction $\text{F}_1$-graphs) is not sufficient, as paths in such hypergraph ignore information about one of vertices corresponding to one of control qubits.

\begin{figure}[t!]
\centering
\includegraphics{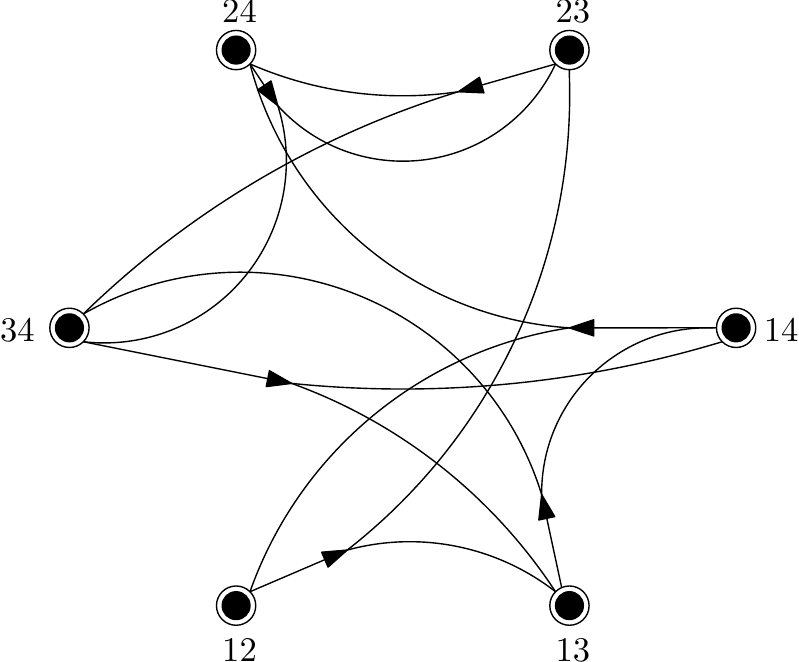}
\caption{An example of an interaction $\text{F}_2$-graph $G$
corresponding to the RUO $\Phi_G^{(3,2)}$. This $\text{F}_2$-graph
belongs to the class of base graphs of RUO $\Phi^{(3,2)}$ acting on four qubits. This graph fulfils the condition given by theorem \ref{theorem_2_1} and it thus belongs to the set of base graphs.}
\label{fig_2_1}
\end{figure}

In the following we define a graph representation of interaction (\ref{def_RUO_2_1}) which takes into account this property. Let us consider an oriented hypergraph $G=(V,E)$, where
$V=\{ij|i<j, j\in\{1,\dots,N\}\}$. Thus vertices of $V$ represent
pairs of qubits, for instance vertex $ij$ represents the pair of qubits
$i$ and $j$. Thus $|V|=\binom{N}{2}$. If $p_{ij,k}>0$, then
the vertices $ij, ik$, and $jk$ are joined with a F-arc
whose tail is the vertex $ij$ and whose head is formed by vertices
$ik$ and $jk$. We call $G$ the interaction $\text{F}_2$-graph
of (\ref{def_RUO_2_1b}). An example of the interaction $\text{F}_2$-graph is depicted in Fig. \ref{fig_2_1}. This representation allows us to
fully describe base $\text{F}_2$-graphs and the base attractor
space by the following theorem, which is given without detailed proof (see appendix).

\begin{theorem}
\label{theorem_2_1}
Base $\text{F}_2$-graphs of the considered
family of interactions are $\text{F}_2$-graphs with the following property.
The $\text{F}_2$ $G=(V,E)$ is a base graph if and only if for each vertex $ij\in V$
and for all $k\in\{1,\dots N\}\backslash\{i,j\}$ exists $l_k\in\{1,\dots N\}$ such that
vertices  $ij$ and $kl_k$ are connected by a path. The basis of common eigenvectors is given by $N+2$ vectors
$\ket{\normalfont{\textbf{0}}_N}$, $\ket{\varphi_N^+}$, and $\{\ket{1_{i,N}}|i\in\{1,\dots,N\}\}$
with $\ket{\normalfont{\textbf{0}}_N}$ and $\ket{\varphi_N^+}$ defined by theorem \ref{theorem_1_2} and
\begin{equation*}
\ket{1_{i,N}}=\ket{0}\otimes\dots\otimes \ket{1_i}\otimes\dots\otimes\ket{0}.
\end{equation*}
All these eigenvectors correspond to the eigenvalue $\alpha=1$ and consequently all base p-attractors solve attractor equations (\ref{eq_attractor_general}) with eigenvalue $\lambda=1$. Together with the identity operator they form the base attractor space for any $N\geq 3$.
\end{theorem}

Equipped with theorem (\ref{theorem_2_1}), we conclude that in the case $N\geq 3$ RUO (\ref{def_RUO_2_1b}) associated with an interaction base $\text{F}_2$-graph generates iterated evolution with the stationary asymptotic state (\ref{eq_evolution_stationary}).
Here, $\mathcal{P}^{(3,2)}$ is the projector onto the $(N+2)$-dimensional space of common eigenvectors.

\subsection{Reaching consensus within qubit networks with cu-interactions}
\label{part_Vc}
From a different perspective quantum networks constitute a well designed model of stochastic distributed quantum processing. In order to achieve a certain goal, individual distant agents of a quantum network interact randomly with their local neighbors according to the prescribed interaction graph. One of basic tasks of distributed processing is reaching consensus among interacting parties. Based on the given situation the desired common consensus can take different forms. Here, in the context of quantum networks, we follow the definition discussed in details in \cite{qnet3}. Consensus is achieved if the total state of the quantum network is permutationally invariant, i.e. reordering of qubits of the network does not lead to a change of its state. Consequently, it induces that all subsystems of the same size are in the same state and thus the expectation value of any local observable provides the same value irrespective which subsystem is being measured. The problem of reaching consensus in quantum networks with swap unitary interaction were successfully resolved in \cite{qnet3,qnet4,qnet5}.

In view of Theorems \ref{theorem_1_2} and \ref{theorem_2_1}, we find out that all attractors of quantum networks with a cu-interaction base graph are permutationally invariant. Therefore, any initial state of such quantum network is asymptotically driven into consensus, i.e. quantum network asymptotically reaches consensus. However, we see that the consensus varies for quantum networks with different cu-interactions. The resulting structure of asymptotic states is determined by a specific information about the initial state which is preserved during the evolution. One can characterize this memory of initial conditions by the set of integrals of motion, i.e. the set of hermitian observables whose expectation values remain constant during the evolution. As the basis of the set of attractors corresponding to eigenvalue $\lambda=1$ can be always chosen hermitian, using Theorems \ref{theorem_1_2} and \ref{theorem_2_1} one can directly construct the whole algebra of integrals of motion of a corresponding quantum network. Moreover, if the interaction topology of the network is given by a base graph the preserved information is uniformly distributed among all subsystems and the consensus is established. An interesting question is whether this information can be extracted by a well designed local measurement solely. The following example illustrates that in some cases it is possible. We assume the observable $X = |\textbf{0}_N\> \<\textbf{0}_N|$, whose expectation value $\Tr [\rho(n) X]$ stays constant during the evolution of any quantum network with cu-interaction. Further, we consider a quantum network with two-qubit cnot-interaction whose qubits interact according to an interaction base graph. We define a local observable acting on one qubit

\begin{figure}[b!]
\includegraphics[width=3.5in]{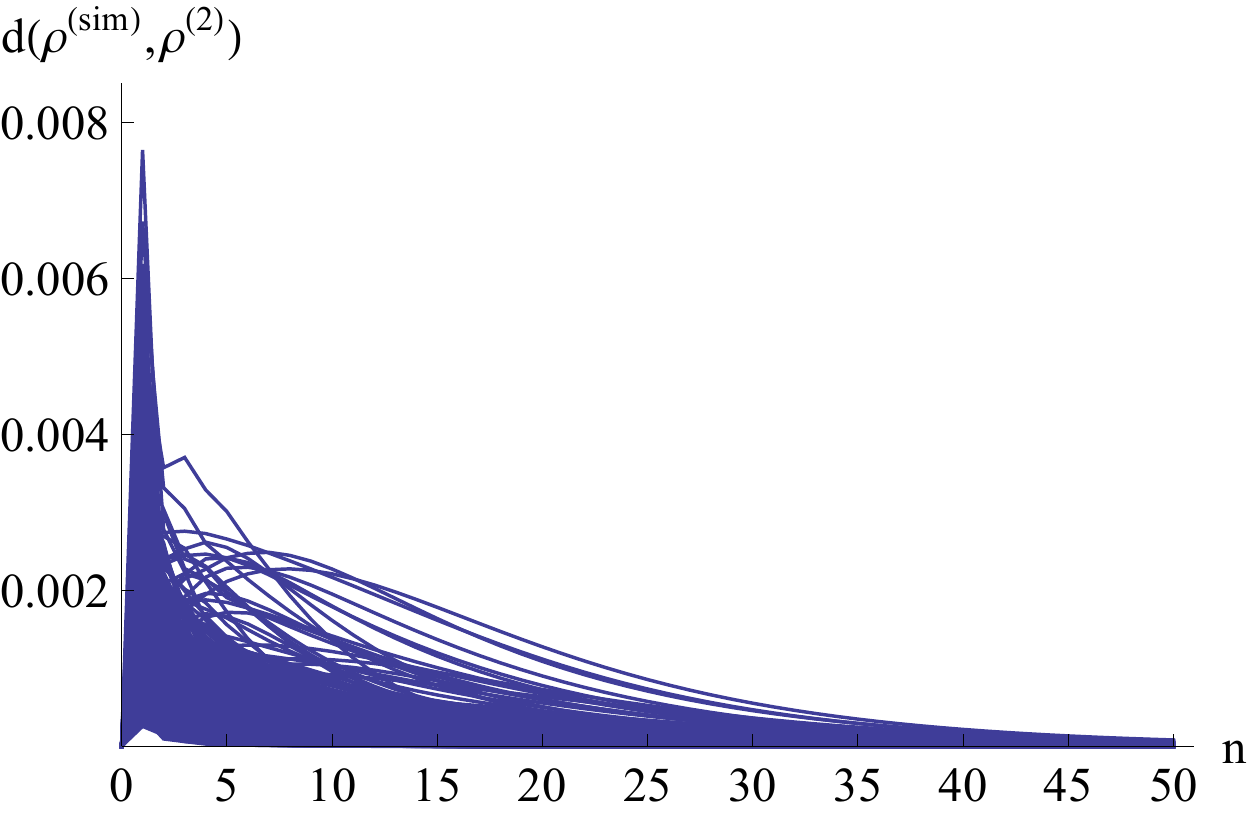}
\caption{Convergence of the state $\rho^{(\text{sim})}(n)$ corresponding to a network consisted of 4 qubits equipped with simultaneous two and three qubit cu-interactions towards the state of the same network equipped with purely two-qubit cu-interactions $\rho^{(2)}$. We plot the distance $d(A,B)=||A-B||_{HS}$ between states $\rho^{(2)}(n)$ and $\rho^{(\text{sim})}(n)$ for an ensemble of 500 randomly chosen initial states, each inserted in operations (\ref{eq_ruo_two_qubit}) and (\ref{phisim}). All relevant graph structures describing the topology of the network belong to corresponding sets of base graphs.}
\label{fig: 23to2}
\end{figure}

\begin{equation}
\sigma = \left( \begin{array}{cc}
1 & 0 \\ 0 & -\frac{2^N-2}{2^N}
\end{array}
\right).
\end{equation}
Using (\ref{eq_asymptotic_evolution_gen}) straightforward algebra reveals equality of expectation values of both observables
\begin{eqnarray}
\Tr \left[\rho(n)X\right] = \Tr\left[\rho_i \sigma \right],
\end{eqnarray}
which applies to any initial state $\rho(0)$, any time $n \in \Naturals$ and any reduced one-qubit density operator $\rho_i$ of the asymptotic state $\rho_{\infty}^{(2)}$. Apparently, mutual two-qubit cnot-interactions uniformly distribute information about the overlap of the initial state $\rho(0)$ with the state $|\textbf{0}_N\>$ among distant constituents of a strongly connected network. This information can be extracted by means of local measurements. We emphasize that if the network is not strongly connected this property is not guaranteed for all initial states $\rho(0)$ any more.

\section{Asymptotic evolution of simultaneous two and three-qubit cu-interaction}
\label{part_VI}
Results of the previous section enable us to study asymptotics of networks with simultaneous two and three-qubit cu-interactions.
In such network we assume that during each iteration, qubits are allowed to undergo either two or three qubit cu-interaction.
As this process is random, one step of evolution is described by the RUO $\Phi^{(\text{sim})}$ defined as

\begin{equation}
\label{phisim}
\Phi^{(\text{sim})}=p^{(2)}\Phi^{(2)}+p^{(3,1)}\Phi^{(3,1)}+p^{(3,2)}\Phi^{(3,2)}
\end{equation}

\noindent with $p^{(2)}+p^{(3,1)}+p^{(3,2)}=1$. Using (\ref{eq_attractor_general}) one can easily deduce that if all
three probabilities $p^{(2)},\ p^{(3,1)}$, and $p^{(3,2)}$ are nonzero, the corresponding attractor space $\text{Atr}(\Phi^{(\text{sim})})$ obeys the relation

\begin{equation*}
\text{Atr}(\Phi^{(\text{sim})})=\text{Atr}(\Phi^{(2)})\cap\text{Atr}(\Phi^{(3,1)})\cap\text{Atr}(\Phi^{(3,2)}).
\end{equation*}

The comparison of base attractor spaces corresponding to three-qubit cu-interactions with
the base attractor space of two-qubit cu-interactions uncovers several
interesting physical implications. Both base attractor
spaces of three-qubit cu-interactions fulfil relations

\begin{figure}[b!]
\includegraphics[width=3.45in]{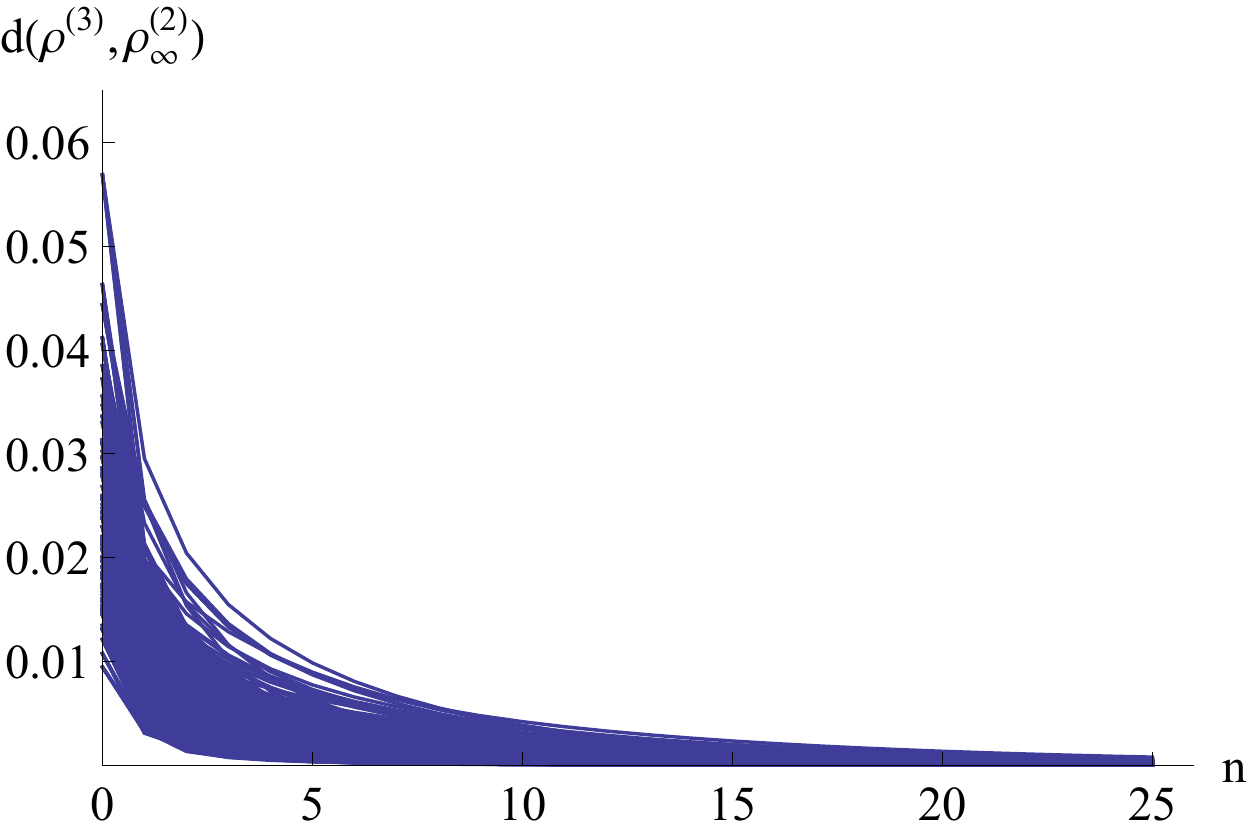}
\caption{Convergence of the state $\rho^{(3)}(n)$ corresponding to a network consisted of 4 qubits equipped with simultaneous three qubit cu-interactions of towards the asymptotic state of the same network equipped purely two-qubit cu-interactions $\rho_{\infty}^{(2)}$. We plot the distance $d(A,B)=||A-B||_{HS}$ between states $\rho^{(3)}(n)$ and $\rho_{\infty}^{(2)}$ for an ensemble of 500 randomly chosen initial states, each inserted in operation (\ref{phisim}) with $p^{(2)}=0$. Both relevant $\text{F}_i$-graphs describing the interaction topology belong to corresponding sets of base graphs.}
\label{fig: 3to2}
\end{figure}

\begin{equation}
\text{Atr}(\Phi^{(2)})\subset\text{Atr}(\Phi^{(3,i)}),\quad i\in\{1,2\},
\end{equation}

\noindent which give us the important attractor space equivalence

\begin{equation*}
\text{Atr}(\Phi^{(\text{sim})})=\text{Atr}(\Phi^{(2)}).
\end{equation*}

\begin{figure}[t!]
\includegraphics[width=3.5in]{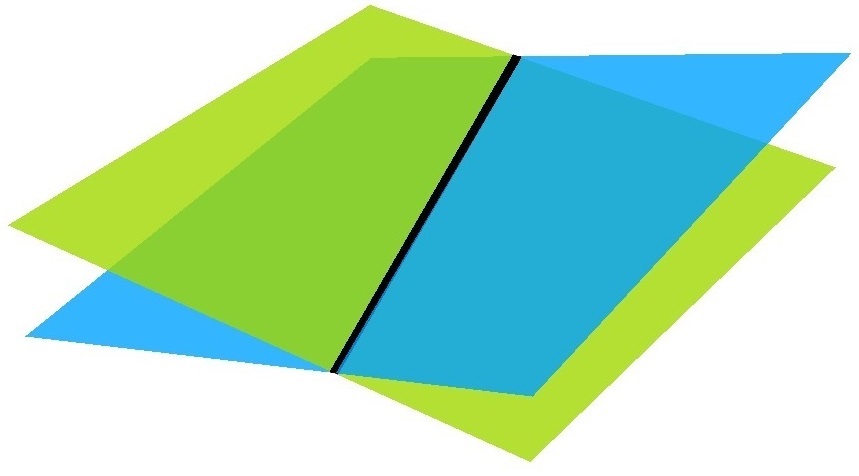}
\caption{Relation between the base attractor space $\text{Atr}(\Phi^{(2)})$
of two-qubit cu-interactions (black line) and base attractor spaces $\text{Atr}(\Phi^{(3,i)})$ of three-qubit
 interactions with $i$ control qubits, $i\in\{1,2\}$ (green and blue plane).}
\label{fig: AtrRel}
\end{figure}

This allows us to identify the privileged role played by two-qubit cu-interactions.
Suppose both two and three qubit cu-interactions are allowed in the given qubit network.
Then, if the interaction graph corresponding to the allowed two-qubit cu-interactions is
strongly connected, the resulting class of possible asymptotic states is not affected by three-qubit
interactions at all. After sufficiently long time we are not able to recognize, whether observed states are results of two-qubit cu-interactions only, or if two-qubit cu-interactions are accompanied also by three-qubit cu-interactions. Figure \ref{fig: 23to2} illustrates this asymptotic behavior for 500 randomly chosen initial states.

Another crucial property is that base attractor spaces of considered three-qubit cu-interactions fulfill the relation
\begin{equation}
\text{Atr}(\Phi^{(3,1)})\cap\text{Atr}(\Phi^{(3,2)})=\text{Atr}(\Phi^{(2)}).
\end{equation}
This relation is displayed schematically in Fig. \ref{fig: AtrRel}. It implies that in the asymptotic regime, three-qubit cu-interactions can be indistinguishable from two-qubit cu-interactions. The state of a qubit network, in which qubits interact exclusively by three-qubit cu-interaction
can converge towards the asymptotic state of a qubit networks with two-qubit cu-interactions. In particular, if both interaction $\text{F}_i$-graphs
belong to corresponding sets of base $\text{F}_i$-graphs, the possible class of asymptotic states is given by
the base attractor space of RUO (\ref{eq_ruo_two_qubit}). An example of this behavior is portrayed for 500 randomly chosen initial states in Fig. \ref{fig: 3to2}.
However as simulations indicate, this condition is not necessary.
Even if one of interaction $\text{F}_i$-graphs does not belong to corresponding sets of base graphs,
the asymptotic dynamics can still tend towards the one of two-qubit cu-interaction for any initial state.

\section{Properties of asymptotic states of three-qubit cu-interactions}
\label{part_VII}
This section provides additional results concerning qubit networks with purely three-qubit cu-interactions.
The main focus is given to properties of three-qubit cu-interactions with one control qubit equipped with a base $\text{F}_1$-graph.

Behavior of qubit networks equipped with interactions (\ref{def_controlled_1_2}) is according to theorem \ref{theorem_1_2} dependent on the value of the parameter $\phi$. Let us consider the parameter
$\phi=\frac{\pi}{2}$. One can easily see, that in this case interactions
(\ref{def_controlled_1_2}) preserve the parity of
$\tau(z)$ (defined in theorem \ref{theorem_1_2}). This results in
the existence of two distinct mixed states $I_{N,E}$ and $I_{N,O}$ belonging to the base
attractor space. Furthermore, for $\phi=\frac{\pi}{2}$ vectors $\ket{\varphi_N^{\pm}}$
have a special form. They can be written as

\begin{align*}
\ket{\varphi_N^+}&=\frac{1}{\sqrt{2^{N}}}\sum_{z\in I(N)}\ket{z},\\
\ket{\varphi_N^-}&=\frac{1}{\sqrt{2^{N}}}\sum_{z\in I(N)}(-1)^{\tau(z)}\ket{z}.
\end{align*}
It is however more convenient to use a different basis. We thus define
\begin{align*}
\ket{\chi_N^E}&=\frac{1}{\sqrt{2^{N-1}-1}}\sum_{\scriptsize{\begin{tabular}{  c  }
  $z\in I(N),$\\
  $\tau(z)=2k,$\\
  $ k\neq 0$\\
\end{tabular}}}\ket{z},\\
\ket{\chi_N^O}&=\frac{1}{\sqrt{2^{N-1}}}\sum_{\scriptsize{\begin{tabular}{  c  }
  $z\in I(N),$\\
$\tau(z)=2k-1$\\
\end{tabular}}}\ket{z}.
\end{align*}
Together with the vector $\ket{\textbf{0}_N}$, these form an orthonormal basis
of the decoherence-free subspace $\mathcal{P}^{(3,1)}\mathscr{H}$. Moreover,
this decoherence-free subspace can be divided into two subspaces with
the help of projectors
$ \mathcal{P}_{E}^{(3,1)}=\ket{\textbf{0}_N}\bra{\textbf{0}_N}+\ket{\chi_N^E}\bra{\chi_N^E}$ and $ \mathcal{P}_{O}^{(3,1)}=\ket{\chi_N^O}\bra{\chi_N^O}$
as $\mathcal{P}^{(3,1)}\mathscr{H}=\mathcal{P}_{E}^{(3,1)}\mathscr{H}\oplus\mathcal{P}_{O}^{(3,1)}\mathscr{H}$.
Let us denote $\tilde{\mathcal{P}}_E^{(3,1)}$  the orthogonal complement to $\mathcal{P}_E^{(3,1)}$ in the subspace of $\mathscr{H}$ with orthonormal basis $\{\ket{z}|z\in I(N), \tau(z)=2k, k\in\mathds{N}\cup\{0\}\}$
and similarly $\tilde{\mathcal{P}}_O^{(3,1)}$  the orthogonal complement to $\mathcal{P}_O^{(3,1)}$ in the subspace of $\mathscr{H}$ with orthonormal basis $\{\ket{z}|z\in I(N), \tau(z)=2k-1, k\in\mathds{N}\}$. The asymptotic state can be then written as

\begin{equation}
\label{As_state_pi/2}
\begin{aligned}
\rho_{\infty}^{(3,1)}=\mathcal{P}^{(3,1)}\rho_{\text{in}} \mathcal{P}^{(3,1)}&+\frac{\text{Tr}\left[\tilde{\mathcal{P}}_E^{(3,1)}\rho(0)\right]}{\text{Tr}\left[\tilde{\mathcal{P}}_E^{(3,1)}\right]}\tilde{\mathcal{P}}_E^{(3,1)}+\\
&+\frac{\text{Tr}\left[\tilde{\mathcal{P}}_O^{(3,1)}\rho(0)\right]}{\text{Tr}\left[\tilde{\mathcal{P}}_O^{(3,1)}\right]}\tilde{\mathcal{P}}_O^{(3,1)}.
\end{aligned}
\end{equation}
The relation (\ref{As_state_pi/2}) is an analogy of the asymptotic
state (\ref{eq_evolution_stationary}). Here, the maximally mixed state living on the orthogonal complement of common eigenvectors is splitted into two by itself invariant maximally mixed states living on subspaces $\tilde{\mathcal{P}}_E^{(3,1)}\mathscr{H}$ and $\tilde{\mathcal{P}}_O^{(3,1)}\mathscr{H}$ respectively.

Another interesting result can be obtained by inspecting the
reduced density matrix corresponding to a single qubit subsystem.
If we define parameters $p_0=\braket{\textbf{0}_N|\rho(0)|\textbf{0}_N}$ and $p_{\pm}=\braket{\varphi_N^{\pm}|\rho(0)|\varphi_N^{\pm}}$, the asymptotic state of a single qubit subsystem of a quantum
network consisting of $N\gg 1$ qubits equipped with
three-qubit cu-interactions with one control qubit for any $\phi\in (0,\pi)$ can be written as

\begin{equation}
\label{singlequbit}
\rho_{\infty}^{(3,1),(1)} =\frac{1}{2}\left( I_1+\vec{a}\cdot\vec{\sigma}\right),
\end{equation}

\noindent with the Bloch vector \cite{schumacher}

\begin{equation*}
\vec{a}=\left((p_+-p_-)\sin\phi,\ 0,\ p_0+(p_+-p_-)\cos\phi\right),
\end{equation*}

\noindent and $\vec{\sigma}=(X,Y,Z)$ being the vector whose elements
are Pauli matrices \cite{nielsen}. Eigenvalues of the
density matrix $\rho_{\infty}^{(3,1),(1)}$ are
$\lambda_{\pm}=\frac{1}{2}(1\pm|\vec{a}|)$.
These eigenvalues thus generally depend on the value
of the parameter $\phi$ with the exception of the case $p_0=0$.

\begin{figure}[t!]
\centering
\includegraphics[width=3in]{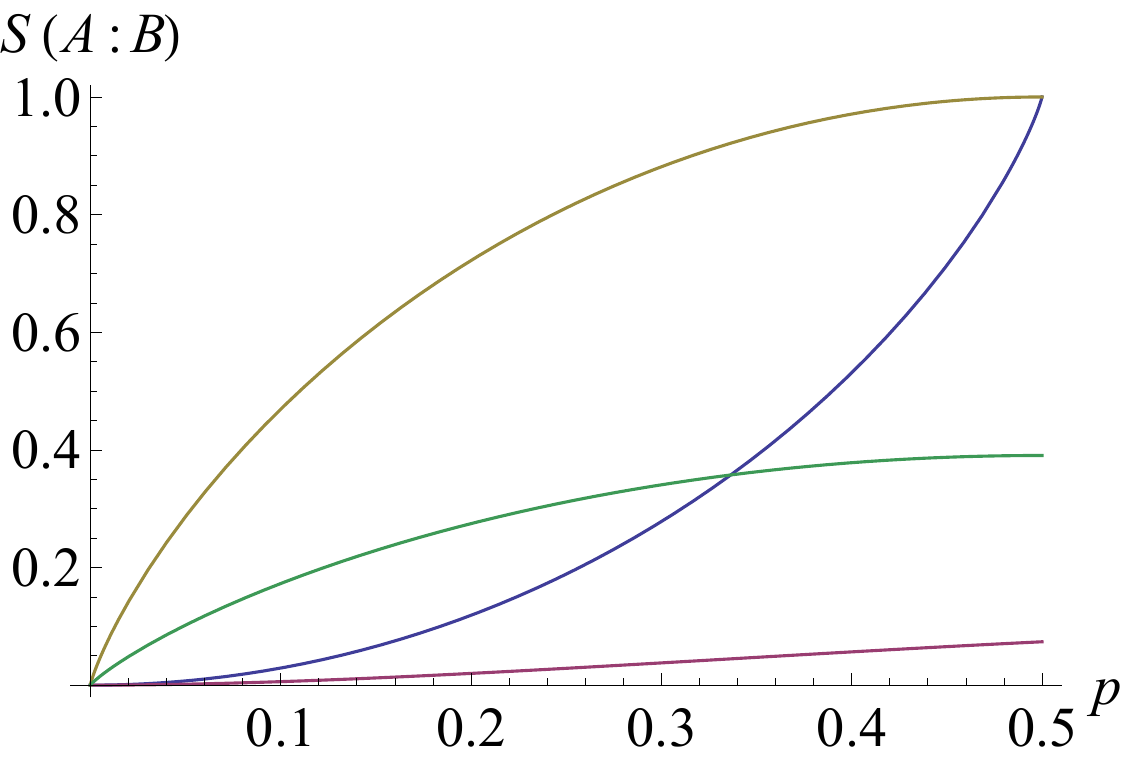}
\caption{The index of correlation of arbitrary two qubits $A$ and $B$ in qubit network interacting with two or three qubit cu-interactions for different cases. The case $p_0=0$, $p_+=p=p_-$ corresponds to the red (blue) line for two-qubit cu-interactions (three-qubit cu-interactions with one control qubit). The case $p_0=1-p$, $p_+=p=p_-$ corresponds to the green (yellow) line for two-qubit interactions (three-qubit cu-interactions with one control qubit).}
\label{correalation}
\end{figure}

Let us now look at entropy related properties of asymptotic states two and three qubit cu-interactions.
A consequence of the inclusion
$ P_N^{(2)}\subset P_N^{(3,i)}$ with $i\in\{1,2\}$ is that
for identical initial states $\rho(0)$ the von Neumann entropy
$S(\rho)=-\text{Tr}[\rho\ \text{ln}\rho]$
\cite{nielsen, barnett, schumacher} of the asymptotic states
$\rho_{\infty}^{(2)}$ and $\rho_{\infty}^{(3,i)}$ satisfy the relation
$S(\rho_{\infty}^{(2)})\geq S(\rho_{\infty}^{(3,i)})$. Concerning three-qubit cu-interactions with one control qubit, there is an additional interesting results concerning
the index of correlation $S(A:B)=S(A)+S(B)-S(A,B)$ \cite{barnett} of two arbitrary qubits $A$ and $B$. The existence of two orthogonal factorizing eigenstates $\ket{\varphi_N^{\pm}}$ results in the possibility of significantly higher correlations between pairs of qubits. This difference between two and three qubit cu-interactions is shown in Fig.\ref{correalation} for $N\gg 1$. The maximal value $S(A:B)=1$ is achieved for three-qubit cu-interactions with one control qubits and for value of corresponding parameters $p_+=p_-=\frac{1}{2}$. This is due to the fact that for $p_+=p_-=\frac{1}{2}$, the single qubit reduced state is according to (\ref{singlequbit}) always the maximally mixed state. Also, for any $\phi\in(0,\pi)$ the eigenvalues of the two qubit reduced state for  $p_+=p_-=\frac{1}{2}$ are $\left\{\frac{1}{2},\frac{1}{2},0,0\right\}$.

\section{Conclusions}
\label{part_VIII}
We investigated the asymptotic dynamics of qubit networks of arbitrary
finite size equipped with two and three-qubit cu-interactions
for a special class of interaction topologies. The main mathematical
results are presented in the form of two theorems (section \ref{part_V}) which characterize
asymptotic states of the considered interaction topologies.
The essential steps of the proof are given in appendix.

For the description of the topology of a given qubit network with
three-qubit cu-interactions we used the concept of hypergraphs.
Although the previously developed formalism \cite{novotny2} of interaction graphs could be
adapted for considered cases by adding coloring of edges,
hypergraphs provide a natural way for the description of
the topology of the interaction. However,
in the case of three-qubit cu-interaction with two control qubits,
there is a significantly different interpretation of vertices of the interaction
graph. This change is necessary for the classification of the network topology.
The described procedure can be extended to general multi-qubit cu-interactions. In this
respect, the given framework is a practical generalization of the
interaction studied previously by Novotn\' y et al. \cite{novotny2}.

Concerning the physics of qubit networks with studied interactions,
results can be summarized in the following way.
First, it is shown in section \ref{part_VI} that enabling three-qubit cu-interactions next to two-qubit
interactions associated with a strongly connected interaction graph
leaves the possible class of asymptotic states unchanged.
This is a particularly important result as it
implies that quantum networks in which qubits interact primarily
via two-qubit cu-interactions mask the effect of qubit
interactions of higher order which thus do not affect the
resulting asymptotic state.

Second, relations between
base attractor spaces of two and three-qubit cu-interactions entail
that in the asymptotic regime, simultaneous action of both types of three-qubit cu-interactions
can be indistinguishable from two-qubit cu-interactions. This result holds for even more general attractor spaces of
three-qubit cu-interactions.

Third, concerning purely three-qubit cu-interactions, results given in section \ref{part_VII} prove
that they have different asymptotic properties compared to properties of controlled two-qubit
interactions. Apart from the above mentioned modification of the graph structure
used for the description of the topology, new attractors emerge in both cases. In the case of controlled three-qubit
interaction with one control qubit, this leads to the existence of two orthogonal factorizing
eigenvectors $\ket{\varphi_N^{\pm}}$. Their presence
in attractor space
is responsible for stronger correlations between pairs of qubits than for
two-qubit cu-interaction. Furthermore, for three-qubit cu-interactions with one control qubit,
one can distinguish two different cases $\phi=\frac{\pi}{2}$ and $\phi\neq\frac{\pi}{2}$
with different asymptotic properties. The characteristic feature of three-qubit cu-interactions
with two control qubits is the explicit dependence of the dimension of the base attractor space
on the size of the quantum network. This property is absent in all other considered
cases. The additional eigenvectors $\ket{1_{i,N}}$, which emerge
in this case, have a simple structure similar to the eigenvector $\ket{\textbf{0}_N}$. Their existence follows directly
from the form of the considered interactions. To summarize,
quantum networks with purely three-qubit cu-interactions
with a fixed number of control qubits have in general a broader class of asymptotic states.

The studied qubit networks exhibit additional interesting features. If each qubit can "see"
all other qubits, all asymptotic states are permutationally invariant. Hence,
a qubit network equipped with a base interaction graph asymptotically establishes consensus among
interacting participants.

Proofs of theorem \ref{theorem_1_2} and theorem \ref{theorem_2_1} indicate the existence of general features
of asymptotic properties of multi-qubit cu-interactions. The number of control qubits
plays a crucial role in the description of networks topology, which can
be done analogously to examined three-qubit cu-interactions.
Increasing the number of target qubits can lead to new properties, e.g. the dependence
of the form of the base attractor space on the value of the parameter $\phi$ or
to the existence of additional attractors, but it does not change the structure
of base graphs, which are formed by all strongly connected interaction $\text{F}$-graphs.

There are several open questions linked to qubit networks with random unitary
operations. Among them the most important question is the
connection between the probability distribution $\{p_i\}$ appearing in (1) and
the rate of convergence of the initial state $\rho(0)$ towards the asymptotic state $\rho_{\infty}(n)$.
Although several numerical calculations were made, therels no clear mathematical result
concerning this dependence. Another group of questions is linked to consensus formation
in qubit networks with different interaction topologies.
These questions will be addressed in future work.
\subsection*{Acknowledgement}
\noindent J. N. and I. J. received funding froms RVO 68407700. I. J. was supported by GACR 13-33906S. J.
M. was supported by SGS13/217/OHK4/3T/14.

\appendix
\section*{Appendix: Some notes about proofs of theorem V.1 and theorem V.2}
The proof of theorem V.1 and theorem V.2 follows concepts developed
in \cite{novotny2}. The purpose of this appendix is to point out
important moments of the proof and main differences with
proof given in \cite{novotny2}. For a detailed proof see \cite{mThesis}.

\begin{figure}[t!]
\centering
\includegraphics[width=3.5in]{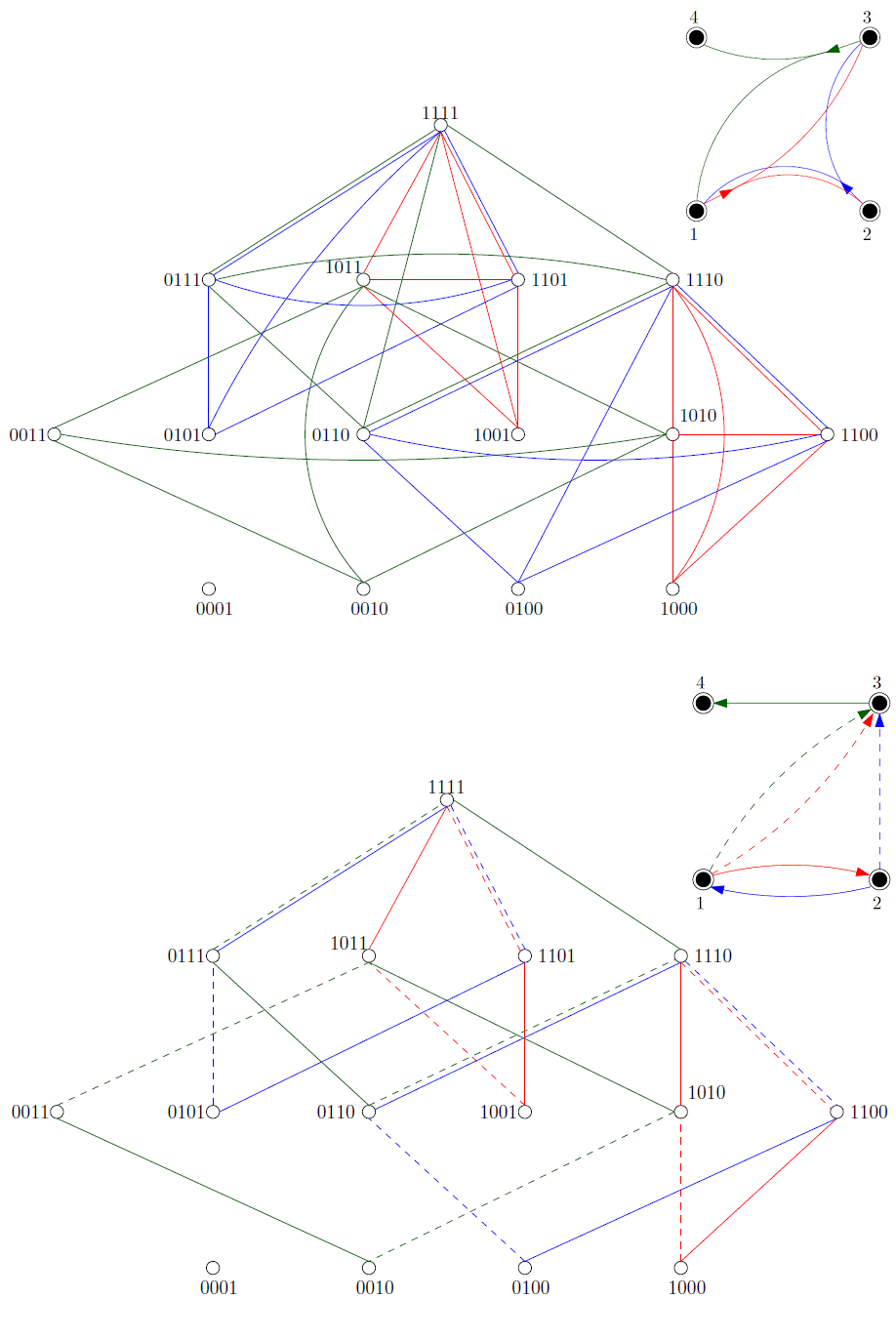}
\caption{The index graph $g^{(3,1)}$ associated with the interaction F-graph $G^{(3,1)}$ and the corresponding index graph $g^{(2)}$.}
\label{fig: IndGraph}
\end{figure}

As the steps of proofs are similar, we start the proof for general three-qubit cu-interaction. The base attractor space definitely appears in the case, where
every possible triplets are present in the set $E$ of an interaction
$\text{F}_i$-graph $G=(V,E)$. Such interaction $\text{F}_i$-graph is called the maximal interaction
$\text{F}_i$-graph. Elements of computational basis are coupled with
each other by operators $U_e$.
For the maximal interaction
$\text{F}$-graph we can divide the pairs of these elements (which
correspond to the elements of density matrix) to the classes $A_i$
with the property, that pairs in each class are all coupled to each
other pairs in the same class and they are not coupled to any pairs
in other classes. To each of these classes corresponds
a single linearly independent attractor which
is then found directly by solving attractor equations. Couplings
between these pairs can be visualized by the so-called index graph
\cite{novotny2}. The index graph is an undirected colored graph,
whose vertices represent elements of computational basis. In its
definition we omit all elements of computational basis which
correspond to the classes $A_j$ which are formed by a single element.
Two vertices $i$ and $j$ of a given index graph are connected
by an edge of color $e$, if the elements $\ket{i}$ and $\ket{j}$
of the computational basis are coupled by the operator $U_e$.

\begin{figure}[t!]
\centering
\includegraphics[width=2in]{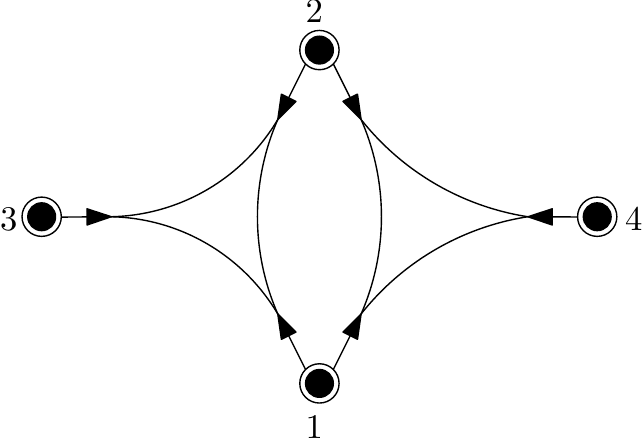}
\caption{The star $\text{F}_1$-graph on 4 vertices. The star $\text{F}_1$-graph is used in the proof of theorem $\ref{theorem_1_2}$ for the case $\phi=\frac{\pi}{2}$.}
\label{fig: stargraph1}
\end{figure}

Now depending on the particular type of interaction we need certain
subgraph of the index graph to be two-connected to obtain the
base $\text{F}_i$-graph. Next, analogously to the procedure used in \cite{novotny2} we prove by induction that we can eliminate
hyperedges from the interaction $\text{F}_i$-graph as long as the newly
emerged interaction $\text{F}_i$-graph fulfils the statement of the corresponding
theorem.

\begin{figure}[b!]
\centering
\includegraphics[width=3in]{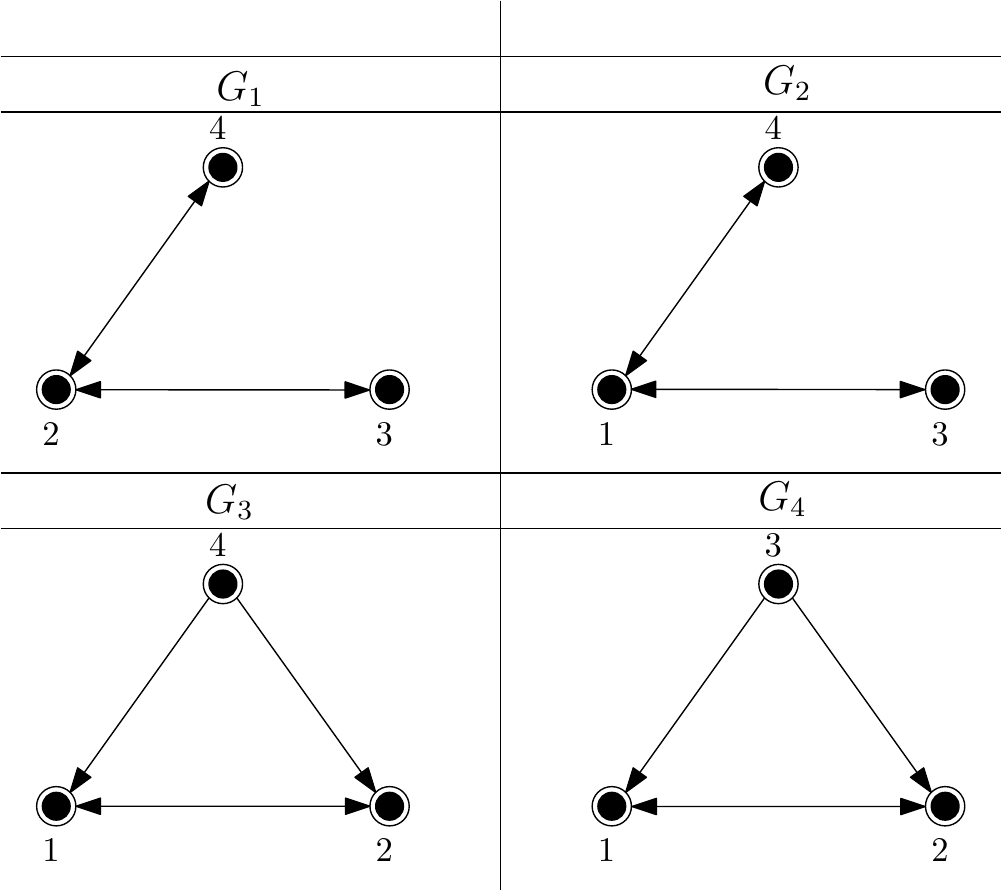}
\caption{The set $\mathcal{G}_4$ corresponding to the star $\text{F}_2$-graph on 4 vertices. The star $\text{F}_2$-graph is used in the proof of theorem $\ref{theorem_2_1}$.}
\label{fig: stargraph2}
\end{figure}

In the case of three-qubit cu-interactions with one
control qubit a single operator $U_e$ couples up to four
elements of the computational basis depending on the particular
element and the value of the parameter $\phi$. For a given index
graph $g^{(3,1)}$ corresponding to the three-qubit cu-interaction with one control
qubit one can find a certain index graph $g^{(2)}$ corresponding to
the two-qubit cu-interaction which is a subgraph of the former index graph
(apart from the case $\phi=\frac{\pi}{2}$). An example of this construction is
displayed in Fig. \ref{fig: IndGraph}.

Afterwards, we prove that one of
these index graphs is two-connected if and only if the other one is
two-connected thus completing the proof for the case
$\phi\neq\frac{\pi}{2}$. For the case $\phi=\frac{\pi}{2}$ the index
graph always has at least two components. We find that the index
graph then corresponds to the base $\text{F}_1$-graph if and only if it has exactly
two components which are two-connected.
We can prove the theorem for this case by taking the generalization of the so-called
star graph (which we call star $\text{F}_1$-graph) \cite{mThesis}. The star $\text{F}_1$-graph is the $\text{F}_1$-graph $G=(V,E)$
with $V=\{1,\dots N\}$ and $E=\{(1,2j)|j\in\{3,\dots N\}\}\bigcup\{(2,1j)|j\in\{3,\dots N\}\}\bigcup\{(j,12)|j\in\{3,\dots N\}\}$. An example
of star $\text{F}_1$-graph on 4 vertices is given in Fig. $\ref{fig: stargraph1}$. Equipped with
this $\text{F}_1$-graph, we follow the same
induction steps as in the proof of the form of base graphs of two-qubit
interactions \cite{novotny2}.

The case of three-qubit cu-interactions with two control
qubits is treated similarly. However, the generalization of the star graph
(which is called star $\text{F}_2$-graph) \cite{mThesis} needs to be constructed in a
different way. First, for an interaction $F_2$-graph $G=(V,E)$
we define a set of ordinary directed graphs
$\mathcal{G}=\{G_i=(V_i,E_i)\}$. Here $G_i$ is an ordinary
directed graph with the set of vertices $V_i=\{j|j\in\{1,\dots N\}\backslash\{i\}\}$
and edges ${(j,k)\in E_i\Leftrightarrow}(i,j;k)\in E$. We define star
$\text{F}_2$-graph as follows: $G=(V,E)$ is a star $\text{F}_2$-graph,
if the corresponding set of graphs $\mathcal{G}$
contains two graphs, which are ordinary star graphs and the
rest of graphs contained in $\mathcal{G}$ are isomorphic
with each other. Figure $\ref{fig: stargraph2}$ shows the set $\mathcal{G}_4$ corresponding to star $\text{F}_2$-graph on 4 vertices. After constructing the generalized star graph the
proof follows the same steps as are used in \cite{novotny2}.

\end{document}